\documentclass[11pt,twoside]{article}
\usepackage{./asp2014}

\aspSuppressVolSlug
\resetcounters

\begin{document}

\title{Cepheid and RR Lyrae Variables as Standard Candles and What Else?}
\author{Anupam Bhardwaj$^1$}
\affil{$^1$Department of Physics \& Astrophysics, University of Delhi, Delhi, India; \email{anupam.bhardwajj@gmail.com}}

\paperauthor{Anupam Bhardwaj}{anupam.bhardwajj@gmail.com}{ORCID_Or_Blank}{University of Delhi}{Department of Physics \& Astrophysics}{New Delhi}{Delhi}{110007}{India}

\begin{abstract}
Classical Cepheid and RR Lyrae variables are fundamental tracers of cosmic distances and stellar evolution and pulsation. Light curve analysis and pulsation properties of these radially pulsating stars provide stringent tests for theoretical evolution and pulsation models. I discuss a detailed light curve analysis of classical Cepheid variables in the Galaxy and the Magellanic Clouds at multiple wavelengths. The variation of light curve parameters as a function of period, wavelength and metallicity is quantified to constrain the theoretical parameters, such as, the mass-luminosity relations obeyed by Cepheids. I briefly summarize the application of multiwavelength light curve data for Type II Cepheid and RR Lyrae variables to study the Milky Way bulge.

\end{abstract}

\section{Introduction}
Classical Cepheid and RR Lyrae variables are primary distance indicators and trace stellar populations of different age and metallicity in their host galaxies. These radially pulsating stars are excellent test beds for the understanding of the theory of stellar pulsation and evolution \citep{cox1980a}. Cepheid variables has been extensively used for extragalactic distance determination as they exhibit a strong Period-Luminosity relation \citep[P-L,][]{leavitt1912}, and to estimate an accurate and precise (upto $\sim2\%$) value of the Hubble constant \citep{freedman2001, riess2016}. RR Lyrae stars, despite being fainter than classical Cepheids, provide an alternate route to cosmic distance scale \citep{beaton2016}. Furthermore, Cepheid and RR Lyrae stars are also used to trace the extinction, morphology and the structural properties of the host galaxy \citep[for example,][]{kunder2008,deb2014, subramanian2015,ripepi2017}. 

Several theoretical studies of Cepheid and RR Lyrae stars has been carried out to predict the observed pulsation properties and the morphology of the light curves of these variables \citep[][and references within]{bono1999b, bono2000, marconi2005, marconi2013, marconi2015}. Most of these studies are based on non-linear convective hydrodynamical pulsation models and can produce light and radial velocity variations for Cepheid and RR Lyrae stars \citep{marconi2013}. The theoretical P-L relations are compared to the observed P-L relations for Cepheids in the Galaxy and the Magellanic Clouds and found to be consistent \citep{bono1999b,caputo2000b,fiorentino2007,bono2010, marconi2013}. Using stellar evolution models, \citet{anderson2016} studied effect of rotation in Cepheid population and showed that Cepheid luminosity increases between the crossing of instability strip and rotation can resolve Cepheid mass discrepancy problem. \citet{smolec2016} used pulsation models to investigate period doubling and dynamical instability in Type II Cepheids for different metallicities. Similarly, \citet{marconi2015} presented new RR Lyrae models for a wide-range of metallicity to study P-L and Wesenheit relations \citep{madore1982} at optical and near-infrared wavelengths.

Recently, \citet{marconi2017} used model-fitting to match observed light and radial velocity curves of fundamental and first-overtone mode Cepheids in the Small Magellanic Cloud (SMC) at optical and near-infrared wavelengths. \citet{bhardwaj2017} carried out a detailed light curve analysis of theoretical models of Cepheids and compared their light curve parameters with observations \citep{bhardwaj2015} to explore constraints for stellar pulsation models. In the following sections, I will summarize the result of these analyses for classical Cepheids and discuss the extension of this work to Type II Cepheid and RR Lyrae stars using data from wide-field variability surveys.  

\section{Light curve analysis of classical Cepheid variables}
\citet{bhardwaj2015} used Fourier decomposition technique to study the light curve structure of Cepheid variables in the Galaxy and the Large Magellanic Cloud (LMC) at optical, near-infrared and mid-infrared wavelengths. The optical light curves of Cepheid variables are taken from the Optical Gravitational Lensing Experiment \citep[OGLE-IV,][]{soszynski2015} while near-infrared light curves are taken from the LMC near-infrared synoptic survey \citep[LMCNISS,][]{macri2015,bhardwaj2016a}. Light curve of Cepheid and RR Lyrae stars display periodic variations so we can fit a Fourier series in the following form: 

\begin{equation}
m = m_{0}+\sum_{k=1}^{N}A_{k} \sin(2 \pi k x + \phi_{k}).
\label{eq:foufit}
\end{equation}

\noindent where, $m$ is the apparent magnitude and $x$ is the pulsation phase. The mean-magnitude ($m_0$) is obtained from the fit and amplitude ($A_k$) and phase ($\phi_k$) coefficients are used to construct Fourier amplitude ratios and phase differences: $R_{k1} = \frac{A_{k}}{A_{1}};~ \phi_{k1} = \phi_{k} - i\phi_{1},~\mathrm{for}~k>1$. The lower order Fourier parameters contain the most characteristic information of the light curves \citep{slee1981}.

\citet{bhardwaj2017} used theoretical models representative of Cepheids in the Galaxy, LMC and SMC at multiple bands and carried out a Fourier analysis to compare the variation of light curve parameters with period, wavelength and metallicity. For a fixed composition, mass-luminosity relations are obtained from stellar evolutionary calculations i.e., canonical relations. In order to account for a possible mass-loss or convective overshooting, a brighter luminosity level (by 0.25 dex) is adopted for each mass i.e., non-canonical relations. Figure~\ref{fig:fig01} displays the variation of theoretical $R_{21}$ parameter with period in $I$-band and the detailed quantitative results for amplitude and phase parameters at multiple wavelengths can be found in \citet{bhardwaj2015, bhardwaj2017}. Figure~\ref{fig:fig01}(a) shows the variation in $R_{21}$ parameter as a function of metallicity. We find a distinct metallicity dependence for short period Cepheid models ($\log(P) < 1$) as $R_{21}$ increase with decrease in metal abundance. Figure~\ref{fig:fig01}(b) shows the variation in $R_{21}$ parameter as a function of stellar mass whereas, Figure~\ref{fig:fig01}(c) displays the same but as a function of temperature. We note that for each mass there are two luminosity levels. In Figure~\ref{fig:fig01}(d), we compare the variation in theoretical $R_{21}$ values with observed $R_{21}$ obtained using $I$-band light curves of Cepheids from OGLE-IV catalogue \citep{soszynski2015}. We find that canonical models display a large offset with respect to observations in the period range, $0.8 < \log(P) < 1.1$. We note that these models have masses greater than $6M_\odot$ and relatively lower temperatures ($5100 \leq T < 5400$ K) in this period range. This suggests that the discrepant models lie closer towards the red edge of the instability strip. 

\articlefigure{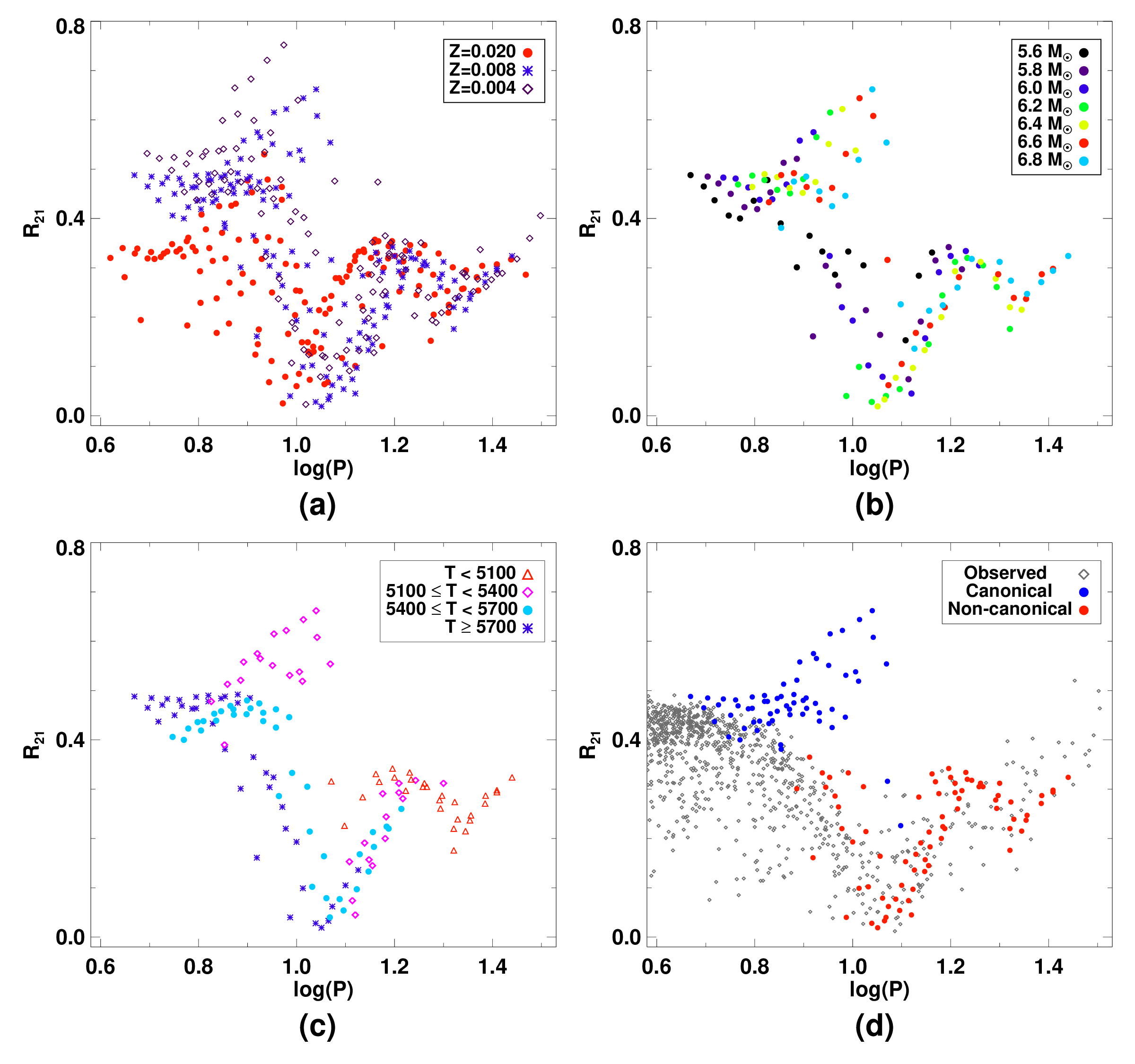}{fig:fig01}{Variation of Fourier amplitude parameter ($R_{21}$) with period as a function of: (a) metal abundance, (b) stellar mass, (c) temperature. In (d), a comparison of observed and theoretical $R_{21}$ values is displayed. Non-canonical models represent a brighter luminosity levels by 0.25 dex for each mass-luminosity relation obtained from stellar evolutionary calculations.}

Fourier amplitude parameters decrease with increase in wavelength at a given period while the phase parameters increase with wavelength \citep[see Figure 6 in][]{bhardwaj2017}. The theoretical peak-to-peak amplitudes at optical bands are greater than the observed amplitudes except for periods close to 10 days. The discrepancy in the amplitude parameters can be decreased by increasing the convective efficiency in the pulsation models. \citet{bhardwaj2017} showed that increasing the mixing length parameter makes the theoretical amplitudes consistent with observed amplitudes in most period bins, however, it also causes a zero-point offset in bolometric mean-magnitudes used to derive P-L relations. It can be seen that an increase in convective efficiency narrows the width of the instability strip, thus, the red-edge becomes bluer \citep{fiorentino2007}.  

\section{Type II Cepheid and RR Lyrae variables from the VVV survey}

Following the analysis on classical Cepheids, \citet{bhardwaj2017a, bhardwaj2017b} studied near-infrared properties of Type II Cepheids in the LMC from LMCNISS and in the Galactic bulge from the VVV survey \citep{minnitivvv2010}. Light curves in $I$ and $K_s$-bands are used to construct templates for Type II Cepheids and the template-fitted mean magnitudes are obtained to derive P-L and Wesenheit relations. The authors find that peak-to-peak amplitudes for W Virginis stars increase as a function of period while these exhibit an opposite behaviour for RV Tauri stars in $IJHK_s$-bands. Fourier parameters in $I$-band for Type II Cepheids in the LMC and the Galactic bulge do not display any significant difference as a function of period. While the $JHK_s$ light curves for Type II Cepheids in the LMC are not as well sampled as the $I$-band light curves from OGLE survey, $K_s$-band photometry of Type II Cepheids in the bulge from VVV survey provides accurate Fourier parameters. \citet{bhardwaj2017b} used Type II Cepheid data from the VVV survey to estimate a distance to the Galactic center and to trace the structure of the Galactic bulge. In an ongoing work (Bhardwaj et. al, in prep.), RR Lyrae light curves from OGLE-IV and VVV survey are being used to study the variation of light curve parameters in $VIK_s$-bands and to trace the old component of the Galactic bulge \citep[for example,][]{dekany2013}. Observed light curve parameters in optical and near-infrared band for Type II Cepheids and RR Lyrae stars will be quantitatively compared with the theoretical models of different metal abundances to complement the studies carried out for classical Cepheids.

\section{Discussion and Conclusions}

I presented a brief overview of the light curve analysis of classical Cepheid variables in order to gain a deeper insight into the physics of stellar pulsation. A comparative analysis of Cepheid light curves provides several important results: 1) Fourier amplitude parameters display clear and distinct variations as a function of period, wavelength and metallicity for both theoretical and observed light curves. 2) Canonical and non-canonical set of models can be differentiated on the Fourier $R_{21}$ plane. 3) Canonical set of models display a large offset with respect to observations for short period Cepheids while non-canonical models are consistent at optical wavelengths. 4) The discrepancy in theoretical amplitude parameters can be resolved by increasing the mixing length parameter. 5) Light curve analysis of Type II Cepheid and RR Lyrae from OGLE-IV and VVV survey can provide additional constraints for the input parameter space of pulsation models and these variables will be used to trace the extinction, metallicity and the structure of the Galactic bulge.

Light curve analysis of Cepheid and RR Lyrae variables can also be used to study period-color and amplitude-color relations at the maximum and minimum light which are used to probe the interaction between stellar photosphere and hydrogen ionization front and to understand the radiation hydrodynamics of the outer envelopes of these variables \citep{smk1993, bhardwaj2014}. In addition, changes in the light curve structure of Cepheid variables can also be correlated with the observed non-linearities in Cepheid P-L relations at the same periods \citep[for example,][]{bhardwaj2016b} and can help to understand the unknown metallicity effects on the P-L relations.

\acknowledgements AB acknowledges support from the Council of Scientific and Industrial Research, New Delhi, India, for a Senior Research Fellowship and thanks Prof. Richard de Grijs for additional support to attend ``Stellar Populations and the Distance Scale'' conference held at the Kavli Institute for Astronomy \& Astrophysics, Peking University, Beijing, China.

\bibliography{thesis}  
\end{document}